# The distribution of local star formation activity as a function of galaxy stellar mass, environment and morphology

E. K. Lofthouse[1]⋆, S. Kaviraj[1], D. J. B. Smith[1] and M. J. Hardcastle[1]

[1]*Centre for Astrophysics Research, School of Physics, Astronomy and Mathematics, University of Hertfordshire, College Lane, Hatfield, AL10 9AB, UK*

4 September 2017

**ABSTRACT**

We present a detailed inventory of star formation in the local Universe, dissecting the cosmic star formation budget as a function of key variables that influence the star formation rate (SFR) of galaxies: stellar mass, local environment and morphology. We use a large homogeneous dataset from the SDSS to first study how the star-formation budget in galaxies with stellar masses greater than $\log(M_*/M_\odot) = 10$ splits as a function of each parameter separately. We then explore how the budget behaves as a simultaneous function of these three parameters. We show that the bulk of the star formation at $z < 0.075$ (∼65 per cent) takes place in spiral galaxies, that reside in the field, and have stellar masses between $10 < \log(M_*/M_\odot) < 10.9$. The ratio of the cosmic star formation budget hosted by galaxies in the field, groups and clusters is 21:3:1. Morphological ellipticals are minority contributors to local star formation. They make a measurable contribution to the star formation budget only at intermediate to high stellar masses, $10.3 < \log(M_*/M_\odot) < 11.2$ (where they begin to dominate by number), and typically in the field, where they contribute up to ∼13 per cent of the total star-formation budget. This inventory of local star formation serves as a $z \sim 0$ baseline which, when combined with similar work at high redshift, will enable us to understand the changes in SFR that have occurred over cosmic time and offers a strong constraint on models of galaxy formation.

## 1 INTRODUCTION

Understanding the processes that drive star formation and the build up of stellar mass is critical to our understanding of galaxy evolution. While it is well-established that the star formation rate (SFR) of the Universe reached a peak at $1 < z < 3$ (Madau et al. 1996; Lilly et al. 1996), there is still significant star formation occurring at the current epoch. A quantitative analysis of exactly where (i.e. in what galaxies) this star formation is taking place is highly desirable, both to understand local stellar mass growth and, more importantly, to establish a $z = 0$ baseline which will enable us to understand the changes in SFR that have occurred over cosmic time.

Several factors are thought to affect the rate at which stars are formed in galaxies, including galaxy stellar mass, morphology and environment. There is a strong correlation between SFR and galaxy stellar mass at all redshifts (Noeske et al. 2007). In particular, in the local Universe, star formation predominantly takes place in relatively massive galaxies, with the largest contribution from galaxies around $\log(M_*/M_\odot) = 11$ (Brinchmann et al. 2004), although the specific SFR (sSFR; the SFR normalised by stellar mass) is known to decrease with stellar mass (Bell et al. 2005; Drory & Alvarez 2008; Damen et al. 2009; Oliver et al. 2010). This change in SFR with stellar mass correlates well with measurements of the local $H_2$ density, which is dominated by massive galaxies with $\log(M_*/M_\odot) > 10$ (Lagos et al. 2014; Keres et al. 2003; Kaviraj et al. 2017). The total gas mass is observed to increase with stellar mass, even though the gas fraction decreases (e.g. Boselli et al. 2001; Genzel et al. 2015). Thus, there is a larger gas reservoir available to form into stars in massive galaxies which translates into a higher overall SFR but the mass normalised sSFR (which should correlate with gas fraction) decreases.

Another factor which can affect the SFR in galaxies is the density of the local environment. The presence of a relationship between average SFR and environment has now been well established, with many works finding strong correlations between the two (e.g. Gómez et al. 2003; Deng 2010; Rasmussen et al. 2012). Both in the local Universe and at higher redshift, we observe that galaxies in higher density environments exhibit, on average, lower star formation rates when compared to those in less dense environments (e.g. Hashimoto et al. 1998; Kauffmann et al. 2004; Tanaka et al. 2004; Scoville et al. 2013; Wetzel et al. 2014). This reduced star formation in groups and clusters can be a result of environmental quenching (Peng et al. 2010; Darvish et al. 2016), with various processes likely to be responsible for driving this reduction in the SFR. For example, in the 'strangulation' scenario a galaxy moving into a higher density environment experiences tidal effects which starve the galaxy of gas. Once the remaining gas has been converted into stars, there is no new supply of gas to replenish the gas reservoir and continue star formation (Larson et al. 1980; Balogh et al. 2000; Balogh & Morris 2000; Peng et al. 2015). Another mechanism is ram pressure stripping, where gas is removed from the galaxy due to interactions with the intracluster medium, as the galaxy moves through the cluster (Gunn & Gott 1972; Quilis et al. 2000). Interactions between galaxies can also remove gas from a galaxy resulting in the shutdown of star formation, a process known as galaxy harassment (Byrd & Valtonen 1990; Moore et al. 1998).

⋆ E-mail: e.k.lofthouse@herts.ac.uk

A third factor which has some bearing on galaxy star formation rates (SFRs) is morphology. At low redshift, star formation occurs mainly in spiral galaxies while ellipticals are more quiescent (Kennicutt 1998; Kaviraj et al. 2007, 2009; Kaviraj 2014b). For example, Kaviraj (2014a) use galaxies drawn from the Sloan Digital Sky Survey (SDSS) Stripe 82 to explore the star formation budget at $z < 0.07$, and find that over half (53 per cent) of the star formation at these epochs is in Sb/Sc galaxies, with only 14 per cent in early-type galaxies.

In order to explore which factors influence the observed evolution of stellar mass growth in galaxies over cosmic time, we need to track where the star formation takes place as a function of key variables such as galaxy morphology, environment and stellar mass. In particular, constructing a detailed inventory of star formation at low redshift is desirable as it gives us a $z = 0$ baseline that can be compared to similar studies over cosmic time, thus providing an empirical measurement of the changes in stellar mass growth over the lifetime of the Universe. Such a measurement also provides a strong statistical constraint on models of galaxy formation.

Previous studies have looked at the local distribution of star formation using samples from the SDSS, such as Gómez et al. (2003), Brinchmann et al. (2004), Kauffmann et al. (2004), Deng et al. (2012a), Kaviraj (2014a), Guglielmo et al. (2015) and Goddard et al. (2017). However, typically this is done as a function of only one or two of the key variables: galaxy morphology, environment or stellar mass rather than a combined approach looking at the distribution of star formation as a function of all key variables simultaneously.

The construction of a comprehensive star formation inventory, as a function of all three variables, using a homogeneous dataset at $z \sim 0$, is the principal aim of this paper. We begin by constructing unimodal distributions of the SFR budget as a function of the individual variables (stellar mass, environment and morphology) and follow this by performing a multi-modal analysis of the star formation budget as a function of all variables simultaneously.

This paper is structured as follows. Section 2 contains a description of the datasets used and a discussion of our sample selection. In Section 3, we look at how the star formation budget depends stellar mass, environment and morphology individually. Section 4 investigates the local SFR budget as a function of all three of these variables simultaneously. This is followed by a summary of our findings in Section 5.

## 2 DATA

We select our galaxy sample from the public seventh data release (DR7; Abazajian et al 2009) of the SDSS. To investigate the properties of star-forming galaxies in the local universe, we restrict the sample to redshifts less than 0.075, yielding 237,649 galaxies. We ensure that the evolution of the SFR function is negligible up to this redshift by repeating our analysis at different redshift ranges up to this limit and find that the results are consistent. To avoid Malmquist bias we construct a volume-limited subsample. Since the SDSS spectroscopic limit in the r-band is 17.77 mag, a volume-limited sample at $z = 0.075$ implies M(r) = −19.8. By choosing to limit our sample to this redshift, we are complete down to a mass limit of $\log(M_*/M_\odot) = 10$. This sample contains 108,255 galaxies and is the final sample used in Section 3.1 to investigate the effect of stellar mass on SFR.

We obtain SFRs for each galaxy from the MPA-JHU catalogue[1]. These are calculated via SDSS spectra and photometry (Brinchmann et al. 2004, hereafter B04). The SFRs are aperture-corrected, to account for the SDSS fibres potentially missing the outer regions of some galaxies at low redshifts using a method based on the procedure described in B04 and Salim et al. (2007). It consists of first calculating the light outside the fibre for each galaxy by subtracting fibre magnitudes from the total photometric magnitudes. Then stochastic models are fitted to the resulting photometry similar to those in Salim et al. (2007).

To account for possible AGN contamination in the SFR values, B04 split the galaxies into subsamples, based on their positions on the BPT diagram (Baldwin et al. 1981). From this, galaxies are classified as purely star-forming, AGN, or composites. For galaxies in the composite and AGN subsamples, they do not use the emission lines and model fits to determine the SFR, as it will likely be affected by the AGN component. Instead, B04 find the SFR by convolving the likelihood distribution of SFR/M for a given D4000 with the likelihood distribution of D4000 for each galaxy.

The total SFR recovered in our volume-limited sample, for all galaxies at $z < 0.075$, is $\log(\mathrm{SFR}/M_\odot \mathrm{yr}^{-1}) = 5.223^{+0.002}_{-0.002}(\mathrm{random})^{+0.01}_{-0.08}(\mathrm{systematic})$. A significant source of error on the SFR measurements used in this work are systematic effects. We use the estimates adopted by B04 for the systematic errors, which takes into account multiple sources of systematic uncertainty. A 2 per cent spread results from the uncertainty in the chosen SFR estimator for non-star-forming BPT classes. Secondly, the aperture corrections are calculated using the average of the likelihood distributions, but using the mode yields values that are up to 15 per cent lower. This is included as a systematic uncertainty on the SFR values. We also include a scatter of 2 per cent due to the differences between using a linear and higher order interpolation scheme to draw values from the likelihood distributions. B04 note that these systematic uncertainties affect the normalisation of the SFR distributions but do not affect the shape. For more details on all sources of systematic error in the SFRs we refer the readers to the discussion in Section 6 of B04. Taken together, this results in a $^{+3}_{-18}$ per cent systematic uncertainty on the SFRs.

---

[1] http://www.mpa-garching.mpg.de/SDSS/DR7/

Using independent estimates of the SFR density in the local Universe we can compare our recovered SFR to the total SFR that would be expected in our volume. We use the measurement of Westra et al. (2010) for $0 < z < 0.1$, corrected for dust by Gunawardhana et al. (2013), of $\rho_* = 6.5 \pm 2.5$ in units of $10^{-3}$ $M_\odot$ yr$^{-1}$ Mpc$^{-3}$ Combining this with the volume covered by our survey gives $\log(\text{SFR}/M_\odot \text{yr}^{-1}) = 5.29 \pm 0.20$, indicating that we recover 87 per cent of the total SFR. If we restrict to only galaxies with $\log(M_*/M_\odot) > 10$, we recover $\log(\text{SFR}/M_\odot \text{yr}^{-1}) = 5.082^{+0.002}_{-0.002}(\text{random})^{+0.02}_{-0.10}(\text{systematic})$. Using the SFRD measurements from B04, we calculate the total SFRD for galaxies with $M_R < -20.2$ which corresponds to our mass limit of $\log(M_*/M_\odot) > 10$. The total expected SFR is then $\log(\text{SFR}/M_\odot \text{yr}^{-1}) = 5.033^{+0.03}_{-0.12}$, and hence we recover 100 per cent of the SFR in the mass ranges studied in this work.

We also employ published stellar masses (Kauffmann et al. 2003) from the MPA-JHU catalogue. These are calculated via fits to the SDSS photometry. We refer readers to Kauffmann et al. (2003) (hereafter, K03) and Salim et al. (2007) for further details on the methodology. In a similar vein to the SFRs, the stellar mass values obtained from the MPA-JHU catalogue could be affected by systematic errors. K03 discuss these sources of uncertainty in Section 6 of their paper. Firstly, aperture effects may affect the measured stellar masses. The K03 results are derived using a 3 arcsecond aperture and then corrected to account for the light outside the fibre. This may result in biases, particularly for spiral galaxies where a significant fraction of the star formation occurs in the outer regions of the galaxy. However, K03 test the effect of aperture bias on their M/L results and find that it has little effect on their final values. Another possible source of systematic uncertainty is the choice of prior. However, by repeating their fitting using different priors, K03 determine that their results are insensitive to the choice of prior. Other possible sources of systematic uncertainty include the stellar population models used which can result in systematic uncertainties comparable to the measurement errors, calibration errors which can lead to systematic offsets in the dust corrections and the choice of IMF. K03 use a Kroupa (2001) IMF, however using other IMFs results in systematic offsets. For example, changing to a Salpeter IMF leads to a factor of 2 increase in stellar mass. Overall, the systematics from star formation histories are likely to be up to ~0.2 dex (Kauffmann et al. 2003; Muzzin et al. 2009). To account for all the systematic errors on the stellar masses we apply a 0.2 dex systematic error to the individual stellar mass values. Typical systematic errors are shown using the single error bars in the top right corner of the left and middle panels in Figures 1, 2 and 3.

To estimate the local environment of each galaxy we use the environment catalogue of Yang et al. (2007), who employ an iterative halo-based group finder to divide the SDSS into separate structures, ranging from isolated galaxies to clusters. For each galaxy in our sample, this catalogue provides an estimate of the mass of the dark matter halo in which it resides and these values are used as a proxy for the environment, following the method of van den Bosch (2002). Galaxies that have dark matter halo masses less than $10^{13} M_\odot$ are classified as being in the field. For a dark matter halo mass estimate of $10^{13} M_\odot$ to $10^{14} M_\odot$, galaxies are classed as being in a group, while cluster galaxies are found in dark matter halos above $10^{14} M_\odot$. Cross-matching to the Yang et al. catalogue yields 103,622 galaxies in the $z < 0.075$ volume-limited galaxy sample with measurements of SFR, stellar mass and local environment data. We use this full sample when investigating the effect of environment in Section 3.2.

Finally, we use the catalogue of bulge-disc decompositions from Lackner & Gunn (2012) for estimates of galaxy morphology. This catalogue uses 2D fits to r-band images of 71,825 SDSS main-sample galaxies, at $z < 0.05$, and 5 different model profiles to calculate bulge-to-total ratios (B/T). The fits include two composite profiles (exponential disc with either an exponential bulge or a de Vaucouleurs bulge), Sérsic profile, exponential disc profile and a de Vaucouleurs profile. For galaxies where the best fit model is an exponential profile, the B/T is set to 0 indicating a pure disc galaxy, while for de Vaucouleurs profiles, galaxies are assigned a B/T value of 1 indicating a pure bulge. Cross-matching to the SDSS $z < 0.075$ volume-limited galaxy sample, yields 21,159 galaxies with SFR, stellar mass and morphological data. Note that the number of galaxies decreases significantly because the Lackner et al. catalogue is restricted to $z < 0.05$. This restricted catalogue is only used when studying the effect of morphology on the SFR budget. To explore how the bulge-disc decompositions from the Lackner et al. catalogue relate to the traditional definitions of morphology, we take a random sample of 100 galaxies from each of the four B/T bins and visually inspect their SDSS images. In the highest B/T bin ($B/T > 0.75$), 92 per cent of galaxies appear to be morphological ellipticals, while in the lowest B/T bin ($B/T < 0.25$) only 4 per cent of galaxies were visually identified as ellipticals (i.e. almost all galaxies are morphological spirals). For the intermediate B/T bins, 79 per cent of galaxies at $0.25 < B/T < 0.5$ are morphological spirals, while at $0.5 < B/T < 0.75$, 66 per cent of galaxies are morphological ellipticals. Therefore, these classifications give us a reasonable way to split our sample of galaxies into traditional morphological classes.

## 3 SFR BUDGET SPLIT BY INDIVIDUAL PARAMETERS

As some of the galaxies in our sample do not have measurements for all three of the variables, we begin by studying the SFR budget as a function of these variables separately. This also allows us to confirm that the trends we find in our analysis are consistent with the wider literature.

### 3.1 SFR budget by stellar mass

We start by investigating how the star formation budget splits by galaxy stellar mass. We divide the full volume-limited sample of 108,255 galaxies into 15 bins of stellar mass, between $10 < \log(M_*/M_\odot) < 11.5$ and calculate the total SFR in each bin. These results are shown by the solid line in the left-hand plot of Fig. 1. The errors are calculated from Poisson errors on the number of galaxies in each bin, combined with the uncertainty on the individual SFRs (excluding systematics) for each galaxy. A typical systematic error is shown in the top right of the plot, calculated from the $^{+3}_{-18}$ per cent systematic errors quoted by B04. The majority of star formation (~63 per cent of the total SFR at $\log(M_*/M_\odot) > 10$) occurs in galaxies of moderate stellar mass where $10.2 < \log(M_*/M_\odot) < 10.8$.

Our results show good agreement with previous observational studies of galaxy stellar mass in the literature. For example, B04 find that star formation predominantly takes place within galaxies around this mass range. We also compare our results to the expected fractions derived from combining the stellar mass function of Weigel et al. (2016) for local late-type galaxies (which contain the majority of local star formation, see Section 3.3) with the star-formation main sequence (Elbaz et al. 2007). This, shown as the dashed line with a confidence interval in grey, is in good agreement with our estimates and is within the expected errors at all stellar masses. The middle panel in Fig. 1 shows how the total stellar mass is distributed between these bins. The solid line shows the stellar mass fraction calculated using stellar masses from the MPA-JHU catalogue (Kauffmann et al. 2003). We compare this to the dashed line which indicates the expected fractions calculated from the stellar mass function for all local galaxies by Weigel et al. (2016) and again find good agreement at all stellar masses.

The contribution to the stellar mass budget peaks at $\log(M_*/M_\odot) = 10.7 - 10.9$, where it contributes ~16.4 per cent of the total SFR over all masses. This range includes the characteristic stellar mass, e.g. $\log(M_*/M_\odot) \sim 10.8$ (Li & White 2009), $\log(M_*/M_\odot) = 10.66 \pm 0.05$ (Baldry et al. 2012). A larger fraction of stellar mass resides in the most massive galaxies compared to the fraction they contribute to the SFR budget, indicating that these galaxies have lower average star formation rates, in line with the findings of previous works (e.g. Bell et al. 2005; Drory & Alvarez 2008; Damen et al. 2009; Oliver et al. 2010; Lara-López et al. 2013). Galaxies with stellar masses below $\log(M_*/M_\odot) = 10.7$ contribute only ~41.7 per cent of the total stellar mass in the local universe and yet contain ~71.2 per cent of the SFR budget, indicating much higher levels of star formation than the typical galaxy. This is summarised in the average sSFR values (right-hand panel of Fig. 1) where the lowest mass galaxies have the highest sSFRs. The average sSFR for each stellar mass bin is found by calculating the sSFR for each galaxy and then taking the average value of the sSFRs in each bin. The error bars shown indicate the standard error on the mean with a typical systematic error in the top right.

### 3.2 SFR budget by environment

We proceed by investigating how the local SFR budget splits by environment, using the sample of 103,622 galaxies that have halo mass estimates. In the left-hand panel of Fig. 2, the SFR budget in each of the different environments defined in Section 2 (Field, Group and Cluster) is split into the same mass ranges as in Section 3.1. For comparison, the stellar mass budget for the same mass bins and again split by environment is shown in the middle panel. The combined results, for all environments, are consistent with the results shown in Fig. 1 (displayed in Fig. 2 as the dashed line). The vast majority of star formation, at $\log(M_*/M_\odot) > 10$, is found in galaxies identified as being in the field (~86.3 per cent) and mostly at lower masses, with ~77.0 per cent hosted by field galaxies with $\log(M_*/M_\odot) < 10.9$.

The relative contribution from groups and clusters increases with stellar mass. For example, in lower mass galaxies ($\log(M_*/M_\odot) < 10.3$), ~90 per cent of star formation is in field galaxies, ~7.8 per cent in groups and only ~2.4 per cent in clusters. However, the contribution from higher density environments (groups and clusters) increases at higher masses, with clusters and groups contributing up to ~48 per cent of the star formation in the highest mass bins.

Similar trends are seen in the stellar-mass budget. Overall, this budget is dominated by field galaxies which contain ~72.3 per cent of the total stellar mass. However, the peak of the stellar mass contribution is at higher masses ($10.7 < \log(M_*/M_\odot) < 10.9$) than the peak contribution to the SFR budget, which is consistent with the stellar-mass-only results in Section 3.1.

While a direct comparison to the wider literature is difficult, because past work has not used these density measures to explore the SFR budget, our results show qualitative consistency with previous studies. For example, many papers such as Balogh et al. (1997); Hashimoto et al. (1998); Gómez et al. (2003); Tanaka et al. (2004) find that there is a decrease in SFRs in higher density environments and fewer galaxies reside in these environments relative to the field resulting in a lower contribution to the overall SFR budget. Deng et al. (2012b) find that the SFRs and sSFRs of SDSS galaxies strongly decrease with increasing density, and that a higher fraction of the stellar mass is hosted by lower mass galaxies in less dense environments compared to high density environments. Our results agree with these works, as we find that at higher stellar masses there is a larger contribution from groups and clusters. For example, ~68 per cent of stellar mass at $\log(M_*/M_\odot) > 11.2$ is in groups and clusters compared to only ~17.5 per cent in these higher density environments at $\log(M_*/M_\odot) < 10.3$.

We complete this section by presenting the average sSFRs for field, group and cluster environments, (right-hand plot in Fig. 2). The average sSFR for each stellar mass bin is found by calculating the sSFR for each galaxy and then taking the average value of the sSFRs in each bin. The error bars shown indicate the standard error on the mean with a typical systematic error shown in the top right. For all environments, there is a trend of decreasing sSFR with stellar mass. Furthermore, the rate at which the sSFR decreases as a function of stellar mass appears to be independent of the local environment of the galaxy.

### 3.3 SFR budget by morphology

We proceed by studying how the SFR budget depends on morphology, using the sample of 21,159 galaxies that have bulge-to-total (B/T) ratio measurements from Lackner et al. We split the sample into four B/T bins: $B/T < 0.25$ (6775 galaxies), $0.25 < B/T < 0.5$ (3604 galaxies), $0.5 < B/T < 0.75$ (2509 galaxies) and $B/T > 0.75$ (8271 galaxies).

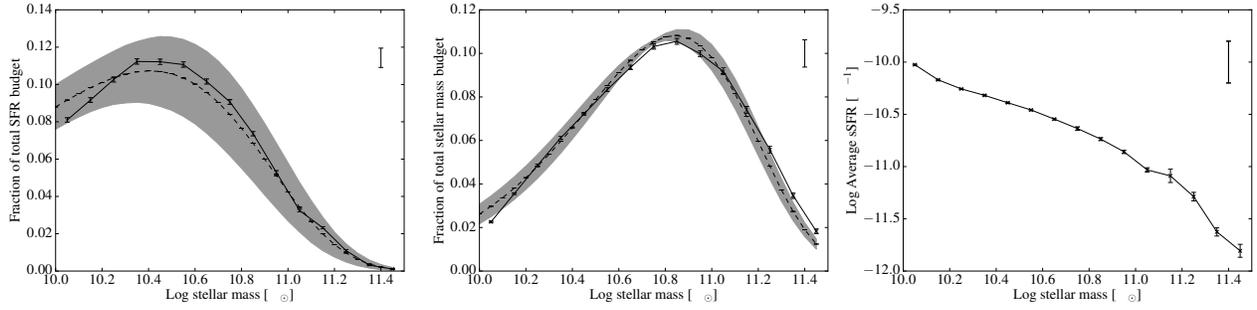

**Figure 1.** Left: The contribution to the total SFR budget as a function of galaxy stellar mass (solid line). For comparison, the expected distribution of star formation, derived from the galaxy main sequence (Elbaz et al. 2007) and the mass function for late-type galaxies (Weigel et al. 2016) is shown by dashed line which is in good agreement with our results. Middle: The contribution to the total stellar mass budget from each stellar mass bin. The dashed line is the expected results derived from the Weigel et al. (2016) mass function for all local galaxies. Right: the average sSFR for galaxies in each stellar mass range. The results in each mass range are plotted at the centre of the bins with errors which are calculated from the standard Poisson errors on the number of galaxies in each bin combined with the uncertainty on the individual SFRs for each galaxy. Typical systematic errors are shown by the error bar in the top right of each plot. This is calculated by including a 0.2 dex systematic error on the stellar masses and a $^{+3}_{-18}$ per cent systematic error for the SFRs.

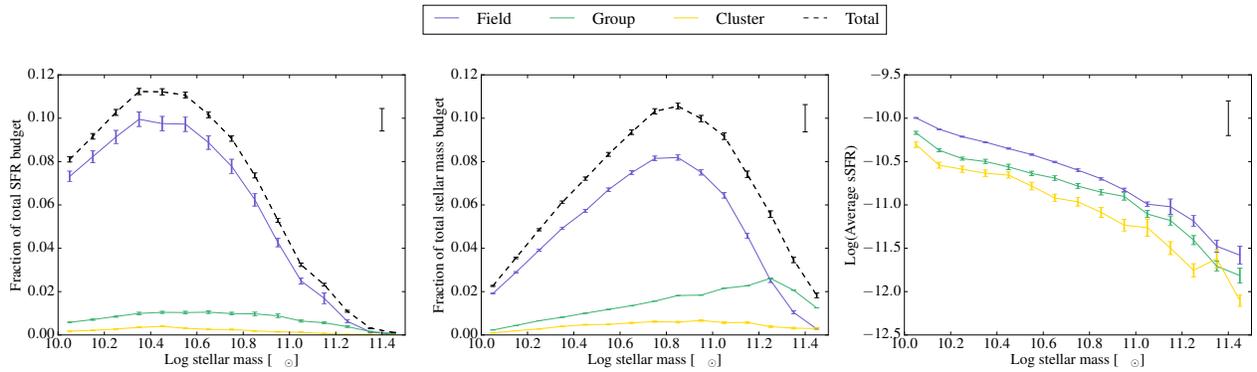

**Figure 2.** Left: the total SFR budget, at $\log(M_*/M_\odot) > 10$, as a function of stellar mass, split into three environments: Field, Groups and Clusters. Middle: the total stellar mass budget for each environment. In both these plots, the dashed lines show the total fractions as calculated for the full sample in Section 3.1. Right: the average sSFR as a function of stellar mass in the three environments. The results in each mass range are plotted at the centre of the bins with errors which are calculated from the standard Poisson errors on the number of galaxies in each bin combined with the uncertainty on the individual SFRs for each galaxy. Typical systematic errors are shown by the error bar in the top right of each plot. This is calculated by including a 0.2 dex systematic error on the stellar masses and a $^{+3}_{-18}$ per cent systematic error for the SFRs.

Fig. 3 (left panel) shows the distribution of the SFR, at $\log(M_*/M_\odot) > 10$, between the four B/T bins and the stellar mass budget for the same B/T ranges (middle). The combined results, for all B/T bins, for the sub-sample used to study the effect of morphology are consistent with the results shown in Fig. 1 (plotted as the dashed line in Fig. 3). The SFR budget is dominated by spiral galaxies, with ∼62 per cent of the total SFR found in galaxies with $B/T < 0.25$ and ∼81 per cent in galaxies with $B/T < 0.5$.

On the other hand, the stellar mass is mostly in bulge-dominated galaxies, e.g. 60.7±0.8 per cent of the total stellar mass is within galaxies with $B/T > 0.75$. This is consistent with previous work on the stellar mass budget in the local Universe (Bernardi 2009; Kaviraj 2014a). For example, Kaviraj (2014a) show that 50.2 ± 0.7 per cent of stellar mass is in early-type galaxies, while ∼70 per cent of stellar mass is in systems with a prominent bulge (early-types and Sa galaxies). It is worth noting that some previous works using local B/T ratios have reported larger fractions of the stellar mass to be in disc galaxies than reported here. For example, Fisher & Drory (2011) find that 75 per cent of the stellar mass is in discs. However, this work was performed with only 320 galaxies and limited to the very local Universe, <11 Mpc, and so is not directly comparable to the results presented here.

The SFR budget shows similar trends, with a decreasing contribution from spirals as mass increases and an increasing contribution from ellipticals. However, even at the highest masses ($\log(M_*/M_\odot) > 11$), the elliptical galaxies never dominate the SFR budget and contribute the same amount of star formation as the spiral galaxies (∼52 per cent compared to ∼48 per cent in spirals). At these higher masses, more galaxies become bulge dominated, therefore an increasing fraction of the budget is expected to be in these systems simply because they will start dominating by number. In comparison, at the lowest masses, ∼79 per cent of the total SFR is in the most disc-dominated systems while only ∼5.0 per cent is in the purely elliptical galaxies.

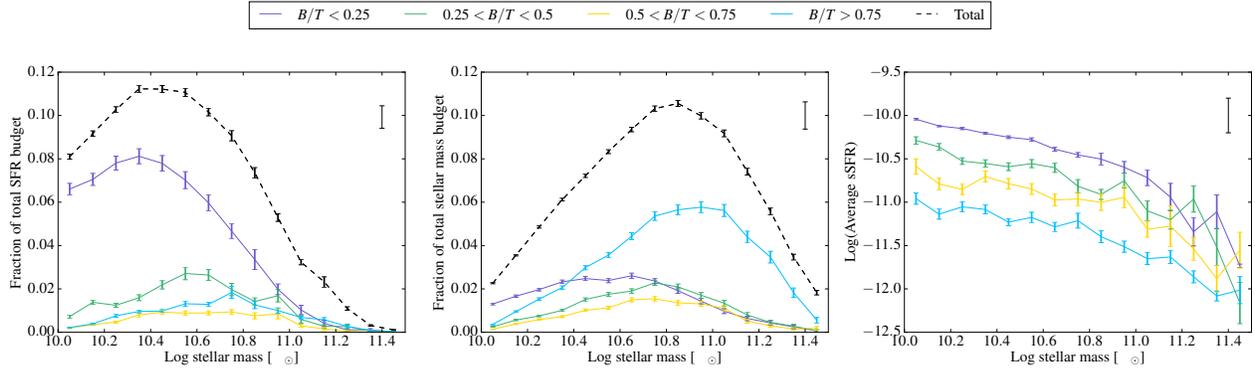

**Figure 3.** The left-hand panel shows the total SFR budget as a function of stellar mass, split into four Bulge-to-Total light ratio bins of B/T< 0.25, 0.25< B/T < 0.5, 0.5< B/T < 0.75, B/T> 0.75. The middle panel shows the contribution to the total stellar mass budget by galaxies in each morphology. In both these plots, the dashed lines show the total fractions as calculated for the full sample in Section 3.1. The right-hand panel shows the average sSFR, as a function of stellar mass, for each morphology. The results in each mass range are plotted at the centre of the bins with errors which are calculated from the standard Poisson errors on the number of galaxies in each bin combined with the uncertainty on the individual SFRs for each galaxy. We show typical systematic errors with the error bar in the top right of each plot. This is calculated by including a 0.2 dex systematic error on the stellar masses and a $^{+3}_{-18}$ per cent systematic error for the SFRs.

Finally, we look at how the average sSFR is affected by morphology. The right-hand plot of Fig. 3 shows the average sSFR in the four B/T ranges. The average sSFR for each stellar mass bin is found by calculating the sSFR for each galaxy and then taking the average value of the sSFRs in each bin. The sSFR by morphology peaks at low stellar masses for all morphologies as seen in the overall results. The rate at which the average sSFR decreases with stellar mass is independent of the morphology of the galaxy. We find that galaxies with a larger bulge have consistently lower sSFRs across all stellar masses. This is in agreement with other studies using alternative measurements of the galaxy morphology (Yuan et al. 2005; Schiminovich et al. 2007; Tzanavaris et al. 2010). For example, Bait et al. (2017) use morphological T-types and find that late-type galaxies exhibit higher sSFRs and that the sSFR is suppressed as the bulge becomes more prominent.

## 4 SFR BUDGET AS A FUNCTION OF STELLAR MASS, MORPHOLOGY AND ENVIRONMENT

In the previous sections, we have established the unimodal trends of SFR with our three variables separately and confirmed that they are consistent with the previous literature. We now study how the star formation budget splits as a function of all three variables simultaneously. This allows us to quantify which combination of these variables hosts the majority of the local star formation and whether the trends in one variable are dependent on the others, e.g. is the dependence of the SFR on stellar mass the same in all environments?

We split the 21,159 galaxies in our sample that have measurements for all three parameters (stellar mass, environment and morphology) into 60 separate bins. These bins include five stellar-mass ranges between $\log(M_*/M_\odot) = 10$ and 11.5, four B/T bins ($B/T < 0.25$, $0.25 < B/T < 0.5$, $0.5 < B/T < 0.75$ and $B/T > 0.75$) and three environments (Field, Group and Cluster).

Fig. 4 shows the percentage of the overall star formation budget (blue) in each of these bins. The four columns indicate different bins in galaxy morphology while the three rows indicate different local environments. The numbers shown in the top right of each panel represent the percentage of the total SFR contained in that panel and the values in the top left show the percentage of stellar mass. Fig. 4 indicates that the majority (∼53 per cent) of local star formation is hosted in the upper left panel i.e. in galaxies that are the most disc-dominated ($B/T < 0.25$) and which reside in field environments. The trend of star formation being predominantly found in lower density environments exists across morphology and stellar masses (e.g. Deng et al. 2012a). In every B/T range, we find that the most star formation is hosted by the field galaxies (top row of Fig. 4). Overall, 85 per cent of the total SFR budget is found in field galaxies with only small contributions from galaxies in groups and clusters. The ratio of the SFR budget in the field, groups and clusters is 21 : 3 : 1.

Looking specifically at galaxy morphology, we find that the majority of the total SFR budget is found in galaxies which have small or non-existent bulges, i.e. a low bulge-to-total light ratio. Galaxies with $B/T < 0.5$, i.e morphological spirals, host ∼81 per cent of the star formation in the local universe. The most disc-dominated galaxies, $B/T < 0.25$, are the largest contributors, hosting ∼61.9 per cent of the SFR budget. Indeed, disc-dominated galaxies host a larger fraction of the star formation budget in all environments from the field to clusters.

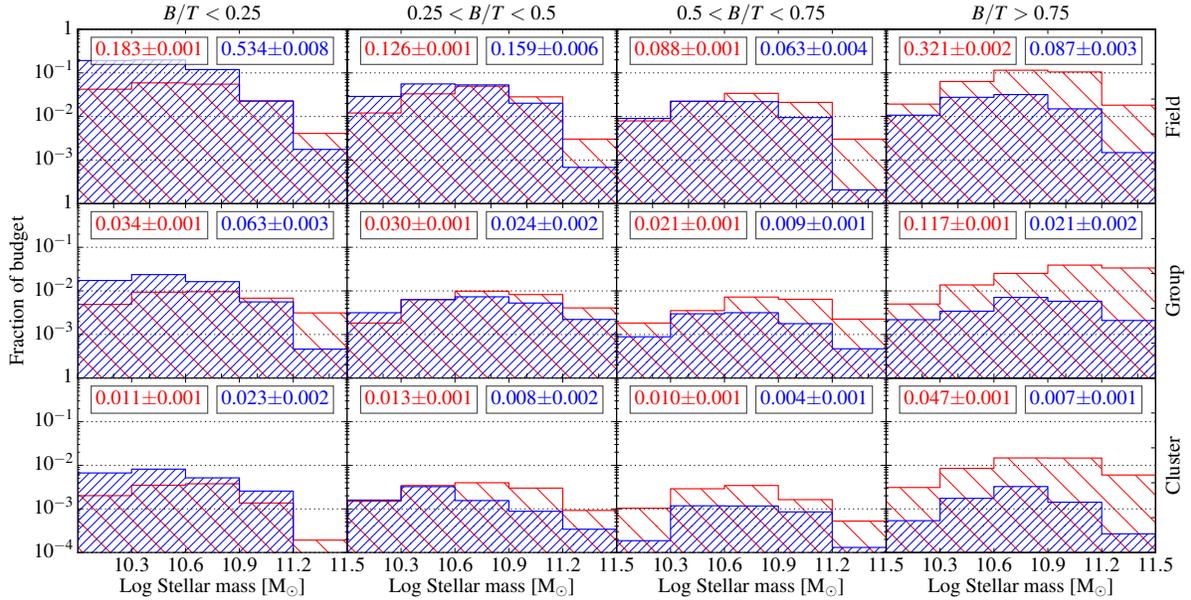

**Figure 4.** The SFR budget (blue) for local galaxies at $z < 0.075$ shown as a function of stellar mass. This has been split into four bins of bulge-total light ratio, indicative of galaxy morphology, along the horizontal axis and simultaneously as a function of environment along the vertical axis. The stellar mass budget is shown in the same bins by the red shaded bars. The numbers shown in the top right of each subplot represent the fraction of the total SFR contained in that panel, while the value in the top left represents the fraction of stellar mass in each panel. The errors shown are the Poisson errors on the number of galaxies in the bin combined with the random errors from the MPA-JHU catalogue.

Looking specifically at stellar mass, we find that, across all environments and morphologies, star formation is preferentially found in systems with low to intermediate stellar masses, with galaxies in the range $\log(M_*/M_\odot) < 10.9$ hosting ∼90 per cent of the SFR in our volume-limited sample. In comparison, using the full volume-limited sample in Section 3.1, we found that ∼87.6 per cent of the SFR budget is below this limit, showing good agreement with the results calculated using galaxies with measurements of all three variables. Across all environments and morphologies, the highest mass range in each panel hosts the smallest fraction of the star formation budget. The overall SFR distribution by stellar mass shows very little variation as a function of environment at constant B/T. However, in all environments we find that the more disc-like galaxies contain a higher contribution from lower masses. For example, at $B/T < 0.25$ 72 per cent of the SFR is at $\log(M_*/M_\odot) < 10.6$, while at $B/T > 0.75$ only ∼40 per cent of the SFR is in this range.

The distribution of local star formation activity across the 60 bins in Fig. 4 could be due to either different average star formation rates in galaxies within each panel or due to the varying number of galaxies, which is reflected in the stellar mass hosted by each bin. To study this further, we also show the stellar mass budget in Fig. 4 using the red bars. In contrast to the SFR budget, this shows that the largest contribution (∼32 per cent) to the local stellar mass budget is in the upper right bin i.e. galaxies that are the most bulge-dominated ($B/T > 0.75$) and reside in field environments. In all environments, and at all stellar masses, the bulge-dominated galaxies ($B/T > 0.5$) host a much larger fraction of the stellar mass budget compared to the star formation budget. This indicates that these elliptical galaxies are less actively star-forming than the disc-dominated galaxies. This is consistent with the results for the full sample shown in Section 3.3, demonstrated by the lower sSFRs for galaxies in the higher B/T bins.

Similarly, we find that higher mass galaxies consistently host a larger fraction of the stellar mass budget than the star formation budget, indicating less active star formation in these galaxies. While galaxies at $\log(M_*/M_\odot) > 10.9$ only host ∼10 per cent of the star formation budget, they host ∼37 per cent of the stellar mass budget in our volume-limited sample.

When studying the effect of the local environment on the stellar mass budget we find a similar trend to the star formation budget. Across all morphologies, the largest contribution to the budgets is from the lower density environments. In every B/T bin, the stellar mass budget is dominated by field galaxies which host ∼78 per cent of the total stellar mass (in agreement with the fraction from the unimodal results of ∼72.3 per cent). This is slightly less than in the star formation budget where field galaxies host over 85 per cent of the local star formation, indicating there are higher average star formation rates in the lowest density bins. The ratio of the stellar mass budget in the field, groups and clusters is 9 : 3 : 1

Combining all aspects of the analysis presented above, we conclude that star formation activity in the local Uni-verse is preferentially hosted by low to intermediate stellar mass galaxies that are disc-dominated and reside in the field. Indeed the specific subset of galaxies which have $B/T < 0.25$, stellar mass less than $\log(M_*/M_\odot) = 10.9$ and reside in the field host ∼51 per cent of the total SFR budget at $z < 0.075$.

## 5 SUMMARY

We have presented a detailed inventory of star formation in the local Universe, using a large homogeneous dataset from the SDSS. Using SFRs and stellar masses from the MPA-JHU catalogue, together with halo mass measurements from Yang et al. (2007) and bulge-to-total ratios from Lackner & Gunn (2012), which are used as proxies for environment and morphology respectively, we have explored the local SFR budget, both as a function of stellar mass, morphology and environment simultaneously and separately for these three variables. Our main results are as follows:

• By studying the SFR budget as a function of the three variables simultaneously, we find that the largest contribution to the SFR budget at $z < 0.075$ is from galaxies which have $B/T < 0.25$, stellar mass less than $\log(M_*/M_\odot) = 10.9$ and reside in the field. The galaxies in this subset host ∼51 per cent of the total SFR budget.

• The SFR budget is dominated by galaxies with masses in the range $10.6 < \log(M_*/M_\odot) < 10.8$ which host ∼63.0 per cent of the SFR in our volume-limited sample.

• The vast majority of star formation (∼65 per cent), at all stellar masses and morphologies, is found in field galaxies, defined as residing in a halo with mass below $\log(M_*/M_\odot) = 13$.

• Galaxies with B/T less than 0.25, which are likely to be morphological spirals, dominate the SFR budget, hosting ∼62 per cent of the total star formation.

• The average sSFR is highest in the lowest stellar mass bin for all environments. The lowest sSFR is found in the highest density environments i.e. clusters. The rate at which the sSFR decreases as a function of stellar mass is independent of local environment.

• The average sSFR by morphology is highest at low stellar masses for all morphologies. Galaxies with higher B/T, i.e. more bulge-dominated, have the lowest sSFR across all stellar masses.

## ACKNOWLEDGEMENTS

SK and MJH are supported by STFC grant ST/M001008/1. SK acknowledges a Senior Research Fellowship from Worcester College Oxford. EKL acknowledges support from the UK's Science and Technology Facilities Council [grant number St/K502029/1].